\documentclass[english,aps,superscriptaddress,reprint,twocolumns]{revtex4-1}
\usepackage[english]{babel}
\usepackage[T1]{fontenc}
\usepackage{amsmath,amssymb,graphicx,mathrsfs,xcolor,babel,textcomp,esint,ulem}
\usepackage[unicode=true,pdfusetitle,bookmarks=true,bookmarksnumbered=false,bookmarksopen=false,breaklinks=true,pdfborder={0 0 0},backref=false,colorlinks=true]{hyperref}

\begin{document}

\title{Dual-Color Laser Induced Terahertz Generation in Strong Field
Approximation}

\affiliation{Shanghai Advanced Research Institute, Chinese Academy of Sciences, Shanghai 201210, China}
\affiliation{Center for Terahertz waves and College of Precision Instrument and Optoelectronics Engineering, Key Laboratory of Opto-electronics Information and Technical Science, Ministry of Education, Tianjin University, China}
\author{Kaixuan Zhang}
\affiliation{Shanghai Advanced Research Institute, Chinese Academy of Sciences, Shanghai 201210, China}
\affiliation{University of Chinese Academy of Sciences, Beijing 100049, China}
\author{Yizhu Zhang}
\email{zhangyz@sari.ac.cn}
\affiliation{Shanghai Advanced Research Institute, Chinese Academy of Sciences, Shanghai 201210, China}
\affiliation{Center for Terahertz waves and College of Precision Instrument and Optoelectronics Engineering, Key Laboratory of Opto-electronics Information and Technical Science, Ministry of Education, Tianjin University, China}
\author{Shuai Li}
\affiliation{Shanghai Advanced Research Institute, Chinese Academy of Sciences, Shanghai 201210, China}
\author{Xincheng Wang}
\affiliation{ShanghaiTech University, Shanghai 201210, China}
\author{Tian-Min Yan}
\email{yantm@sari.ac.cn}
\affiliation{Shanghai Advanced Research Institute, Chinese Academy of Sciences, Shanghai 201210, China}
\author{Y. H. Jiang}
\email{jiangyh@sari.ac.cn}
\affiliation{Shanghai Advanced Research Institute, Chinese Academy of Sciences, Shanghai 201210, China}
\affiliation{University of Chinese Academy of Sciences, Beijing 100049, China}
\affiliation{ShanghaiTech University, Shanghai 201210, China}

\begin{abstract}
  The mechanism of the terahertz (THz) wave generation (TWG) in dual-color
  fields is elucidated within the theoretical framework of single-atom based
  strong field approximation (SFA). Evaluating the transition dipole moment,
  the continuum-continuum (CC) transition, rather than the continuum-bound
  recombination for the high-order harmonic generation, is confirmed to be the
  core mechanism of the TWG. The analytic form of the SFA-based CC description
  is consistent with the classical photoelectric current model, establishing
  the quantum-classical correspondence for the TWG. The theory is supported by
  parametric dependence of experimental THz yields calibrated by the joint
  measurement of the third-order harmonics. Present studies leave open the
  possibility of probing the ultrafast dynamics of continuum electron.
\end{abstract}
\maketitle

Terahertz (THz) wave generation (TWG) using dual-color femtosecond pulse,
typically focusing 800 nm and 400 nm beams into gas-phase medium, allows for
the convenient and efficient access to moderately strong ultra-broadband THz
pulse {\cite{Cook2000}}. Although the approach is widely applied in several
disciplines, the underlying generation mechanism is still under discussion
{\cite{PRL2016Andreeva,Zhangprl2017}}. The interpretations so far proposed to
unravel the mechanism of TWG, e.g., the four-wave mixing (FWM) and the
photocurrent (PC) models, are rather distinctive in their appearances, and the
intrinsic physical pictures are completely different. The FWM, as in crystal
nonlinear optics {\cite{Cook2000}}, explains the nonlinear THz emission based
on the quantum perturbation theory, whereas the PC model starting with the
plasma formulates the emission process classically {\cite{Kim2007,Kim2008}}.

Although the laser plasma is considered the source of the TWG since the first
observation from the laser-gas interaction {\cite{Hamster1993PRL}}, it is
still unclear whether the plasma effect is a must ingredient. Similar debate
arose in the early days when high-harmonic generation (HHG) was studied.
Nowadays it is widely accepted that the HHG is a nonperturbative strong field
process dominated by the continuum-bound transition within a single atom or
molecule, i.e., recombination of released electron with its parent ion after
the ionization. A question naturally arises whether the TWG mechanism can also
be clarified by the established strong field theory without the necessity of
calling upon plasma effects. If so, the unified theory for both the TWG and
the HHG would provide the complementary description of the ionization
dynamics, the possible detection scheme and potential applications. In fact,
the TWG has been numerically studied by solving time-dependent Schr{\"o}dinger
equation {\cite{Karpowiczprl2009,Zhangprl2012,Kostinprl2016}}, accounting for
the strong field dynamics of a single atom. Besides, the strong field
approximation (SFA), which has been extensively developed to treat various
strong field phenomena, including above threshold ionization (ATI), high-order
ATI, multiple-ionization, and HHG {\itshape{et al.}}, was also applied to the
TWG {\cite{Zhoupra2009,BalciunasOE2015}}. However, the link among various
theories is still unclear, and the mechanism of the TWG requires more
investigation.

In this letter, the origin of the TWG in dual-color fields is inspected by
deriving the transition dipole moment under the SFA. The evidences about the
dependence of the THz signal on delay-phase and relative polarization angles
are presented with the accompanied experiment. The delay dependence between
the TWG and third harmonic generation (THG) confirms that the TWG is dominated
by the continuum-continuum (CC) transition, rather than the continuum-bound
(CB) recollision. Our work has manifold implications. From theoretical aspect,
the application scenario of the SFA is further expanded, bringing the TWG
explicable under the framework of the strong field physics similar to the HHG.
From application aspect, it implies that the TWG is still obtainable through
the CC transition even when the neutral atoms are fully depleted by the strong
pump laser, showing the possibility to achieve intense THz fields by pumping
gas-phase medium with extremely strong laser. Moreover, since the TWG is
encoded by the time-dependent information of the continuum electron, it can be
used as a spatial-temporal probe in microscopic scale, complementary to HHG
spectral lineshape and photoelectron momentum distributions, to trace
ultrafast dynamics of continuum electron in atoms and molecules
{\cite{Babushkin_arXiv}}.

Quantum mechanically, the radiation is induced by the time variant dipole
moment $\ensuremath{\boldsymbol{d}} (t) = \langle \Psi (t) |
\hat{\ensuremath{\boldsymbol{r}}} | \Psi (t) \rangle = \langle \Psi_0 |
\hat{U} (t_0, t)  \hat{\ensuremath{\boldsymbol{r}}}  \hat{U} (t, t_0) | \Psi_0
\rangle$ with the time evolution operator $\hat{U}$ and the initial wave
function $| \Psi_0 \rangle$. Using the Dyson series for $\hat{U}$, it is shown
$\ensuremath{\boldsymbol{d}} (t) =\ensuremath{\boldsymbol{d}}^{(0)} (t)
+\ensuremath{\boldsymbol{d}}^{(1)} (t) +\ensuremath{\boldsymbol{d}}^{(2)} (t)$
including three components {\cite{becker_unified_1997}}. The first one
$\ensuremath{\boldsymbol{d}}^{(0)} (t) = - \langle \Psi_0 (t) |
\hat{\ensuremath{\boldsymbol{r}}} | \Psi_0 (t) \rangle$ vanishes in the
spherically symmetric system. The second term
$\ensuremath{\boldsymbol{d}}^{(1)} (t) = \mathrm{i} \int_{t_i}^t \mathrm{d} t'
\langle \Psi_0 (t) | \hat{\ensuremath{\boldsymbol{r}}}  \hat{U} (t, t')
\hat{W} (t') | \Psi_0 (t') \rangle +$c.c. describes the transition between CB
states. The last term $\ensuremath{\boldsymbol{d}}^{(2)} (t) = - \int_{t_i}^t
\mathrm{d} t''  \int_{t_i}^t \mathrm{d} t'  \langle \Psi_0 (t'') | \hat{W}
(t'') \hat{U} (t'', t) \hat{\ensuremath{\boldsymbol{r}}} \hat{U} (t, t')
\hat{W} (t') | \Psi_0 (t') \rangle$ takes the form of CC transition. Since the
external light field is intense, the situation enters the scope of the strong
field physics and a natural choice to tackle with the problem is the SFA
theory. Essentially, the SFA neglects the influence from the Coulomb potential
of the ionic core. Hence, $\hat{U}$ can be substituted by $\hat{U}^{(V)}$, the
evolution operator of the Volkov state which is the eigenstate of an electron
in the external light field alone, to simplify the further derivation. With
the SFA, \ $\ensuremath{\boldsymbol{d}}^{(1)} (t)$ is used to describe the
HHG, which is essentially the widely used Lewenstein's model of an
illustrative interpretation: the atomic ionization is followed by the
transition of the continuum electron back to the bound state, more
intuitively, the recollision of the released electron to its parent core,
yielding the HHG. The contribution of $\ensuremath{\boldsymbol{d}}^{(2)} (t)$
to the HHG is usually negligible {\cite{becker_unified_1997}}, as only the
"hard" recollision leads to photons of high energy, whereas
$\ensuremath{\boldsymbol{d}}^{(2)} (t)$ is the "soft" transition between
continuum states and the energy of the radiation photons is expected to be
small. For the THz photons with small energy, the contribution from
$\ensuremath{\boldsymbol{d}}^{(2)} (t)$ should be considered, though it is
rarely mentioned {\cite{Zhoupra2009,BalciunasOE2015}}. In this work, the TWG
mechanism is investigated based on the analysis of
$\ensuremath{\boldsymbol{d}}^{(2)} (t)$, which is referred to as the SFA based
CC transition (SFA-CC).

With the radiation given by the acceleration form
$\ddot{\ensuremath{\boldsymbol{d}}} (t)$, we evaluate the emission field from
the derived $\ensuremath{\boldsymbol{d}}^{(2)} (t)$ (see Supplementary S1 for
details),
\begin{equation}
  \ensuremath{\boldsymbol{\mathcal{E}}} (t) \propto
  \ddot{\ensuremath{\boldsymbol{d}}}^{(2)} (t) \equiv
  \ensuremath{\boldsymbol{a}}_1 (t) +\ensuremath{\boldsymbol{a}}_2 (t) .
  \label{eq:Et-THz-CC}
\end{equation}
The first term
\begin{equation}
  \ensuremath{\boldsymbol{a}}_1 (t) =\ensuremath{\boldsymbol{E}} (t) 
  \int_{t_i}^t \mathrm{d} t''  \int_{t_i}^t \mathrm{d} t' \eta (t', t'') W
  (t') W^{\ast} (t'') \mathrm{e}^{\mathrm{i} S_{\ensuremath{\boldsymbol{k}}',
  I_{\text{p}}} (t', t'')}, \label{eq:a1}
\end{equation}
where $\eta (t', t'') = \left( \frac{2 \pi}{\mathrm{i} (t' - t'')} \right)^{3
/ 2}$ depicts the diffusion of the electronic wave packet, $W (t')
=\ensuremath{\boldsymbol{\mu}} [\ensuremath{\boldsymbol{k}}'
+\ensuremath{\boldsymbol{A}}(t')] \cdot \ensuremath{\boldsymbol{E}} (t')$ is
the interaction of the electron with the incident light field
$\ensuremath{\boldsymbol{E}} (t')$. Here $\ensuremath{\boldsymbol{k}}'
=\ensuremath{\boldsymbol{k}}' (t', t'') \equiv -
[\ensuremath{\boldsymbol{\alpha}}(t') -\ensuremath{\boldsymbol{\alpha}}(t'')]
/ (t' - t'')$, and $\ensuremath{\boldsymbol{\alpha}} (t) = \int^t \mathrm{d}
t' \ensuremath{\boldsymbol{A}} (t')$ is the excursion of the electron and
$\ensuremath{\boldsymbol{A}} (t')$ is the vector potential. The ionization
rate is related to $S_{\ensuremath{\boldsymbol{k}}', I_{\text{p}}} (t', t'') =
\int_{t''}^{t'} \mathrm{d} t'''  \left\{ \frac{1}{2}
(\ensuremath{\boldsymbol{k}}+\ensuremath{\boldsymbol{A}}(t'''))^2 +
I_{\text{p}} \right\}$ with $I_{\text{p}}$ the ionization energy. The presence
of both $W (t')$ and $W (t'')$ indicates the two electronic continuum states
are involved, and $\mathrm{e}^{\mathrm{i} S_{\ensuremath{\boldsymbol{k}}',
I_{\text{p}}} (t', t'')}$ indicates the joint occurrences of these continuum
states created by ionization. The second term of Eq. (\ref{eq:Et-THz-CC})
emerges when the emission time $t$ approaches the ionization time $t'$,
\begin{equation}
  \ensuremath{\boldsymbol{a}}_2 (t) = - 2 \text{Re} \int_{t_i}^t \mathrm{d} t'
  \eta (t, t') W (t) W^{\ast} (t')  [\ensuremath{\boldsymbol{k}}'
  +\ensuremath{\boldsymbol{A}}(t)] \mathrm{e}^{\mathrm{i}
  S_{\ensuremath{\boldsymbol{k}}', I_p} (t, t')}, \label{eq:a2}
\end{equation}
which is referred to as the temporal boundary term. Here,
$\ensuremath{\boldsymbol{k}}' =\ensuremath{\boldsymbol{k}}' (t, t')$.

\begin{figure}[h]
  \includegraphics[scale=1]{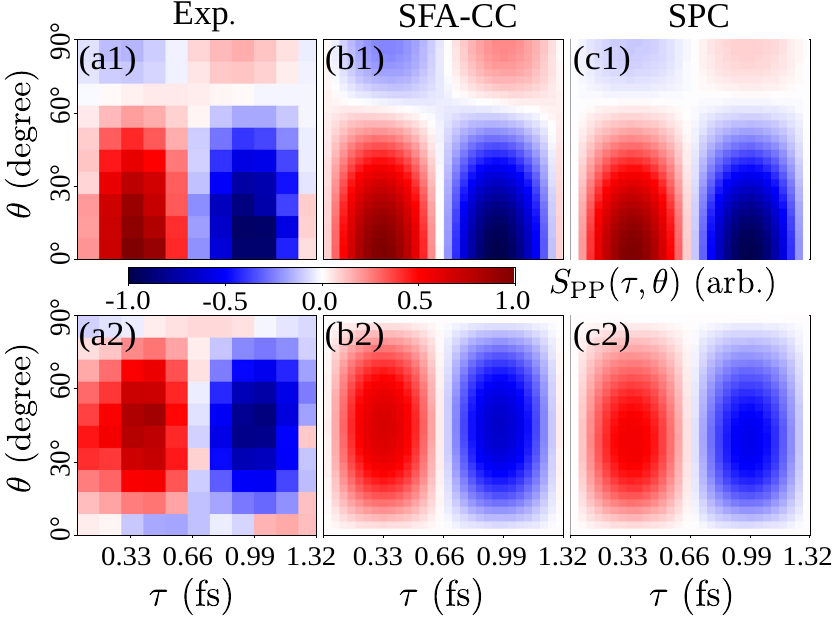}
  \caption{Comparison of PP distributions, $S_{\text{THz}, s} (\tau, \theta)$
  (upper row) and $S_{\text{THz}, p} (\tau, \theta)$ (lower row),
  respectively, for the $s$- and $p$-components of
  $\ensuremath{\boldsymbol{\mathcal{E}}}_{\text{THz}} (t)$ obtained from
  experiment (a) and theoretical models of the SFA-CC (b) and the SPC
  (c).\label{fig:pp-comparison}}
\end{figure}

The TWG in dual-color fields is determined by Eqs.
(\ref{eq:Et-THz-CC})-(\ref{eq:a2}). We apply these equations to examine the
delay- and polarization-dependence of the TWG, and the results are compared
with our experiment. Here, the femtosecond laser with pulse energy of
$\sim$1.75 mJ and duration of $\sim$35 fs passes through a BBO crystal,
generating 800/400 nm two-color laser fields with a intensity ratio of 3:1.
The two-color laser is then focused by a reflection mirror of 100 mm focal
length to ionize the atmospheric air, and the tight focus scheme is adapted so
that the propagation effect in plasma is negligible. Throughout the
measurement, the $2 \omega$ wave is kept {\itshape{s}}-polarized, while the
relative polarization angle $\theta$ and time delay $\tau$ between the
two-color fields can be independently controlled. The vector of the emitted
THz electric field,
$\ensuremath{\boldsymbol{\mathcal{E}}}_{\ensuremath{\operatorname{THz}}}
(\tau, \theta)$, is recorded with the electro-optic sampling technique. More
experimental details and the definition of observables can be referred from
the Supplementary S2. The components of peak-peak (PP) values along the
orthogonal polarizations, $S_{\text{THz}, s}  (\tau, \theta)$ and
$S_{\text{THz}, p}  (\tau, \theta)$ extracted from
$\ensuremath{\boldsymbol{\mathcal{E}}}_{\ensuremath{\operatorname{THz}}}
(\tau, \theta)$, are presented in Fig. \ref{fig:pp-comparison}(a).

Since it is nontrivial to precisely acquisite the time delay $\tau$ between
the two-color fields, a joint measurement of the intensities of the THG,
$I_{\text{3rd}} (\tau, \theta)$, is performed. We notice that the THG
emissions evaluated with all theoretical models present the similar patterns
that the maximum of $I_{\text{3rd}} (\tau, \theta)$ appears at $\tau = 0$
(Supplementary S3). Hence, the time delay zero of the $\tau$-dependent signals
can be reliably determined by locating the maximum of the $I_{\text{3rd}}
(\tau, \theta)$.

In our measurement, the $\tau$-dependent distributions $S_{\text{THz}, s (p)}
(\tau)$ [Fig. \ref{fig:pp-comparison}(a)] and $I_{\text{3rd}} (\tau)$
(Supplementary S3) show the antiphase relationship that the maximum TWG along
$\tau$ coincides with the minimum THG. On one hand, the distribution in
antiphase clearly rules out the contribution from CB transition
$\ddot{\ensuremath{\boldsymbol{d}}}^{(1)} (t)$, since it contradicts the
synchronized distributions of $S_{\text{THz}, s (p)} (\tau)$ and
$I_{\text{3rd}} (\tau)$ as predicted by
$\ddot{\ensuremath{\boldsymbol{d}}}^{(1)} (t)$ (see results in Supplementary
S4). On the other hand, the PP values from the CC transition
$\ddot{\ensuremath{\boldsymbol{d}}}^{(2)} (t)$, as shown in Fig.
\ref{fig:pp-comparison}(b), well reproduces the salient experimental
characteristics, e.g., the decrease and revival of $S_{\text{THz}, s} (\tau,
\theta)$ when $\theta$ increasing, confirming the role of the single atom
ionization process in the TWG.

The distribution is also evaluated with the PC model, where the TWG is
determined by the time-variant plasma density,
$\ensuremath{\boldsymbol{\mathcal{E}}}_{\mathrm{THz}} (t) \propto \partial
\ensuremath{\boldsymbol{j}} (t) / \partial t = e^2 N (t)
\ensuremath{\boldsymbol{E}} (t) / m$. The transient electron density $N (t)$
originates from the accumulated electrons from ionization, satisfying
\begin{equation}
  \partial_t N (t) = [N_g - N (t)] w (t), \label{eq:PC-Nt}
\end{equation}
where $N_g$ is the initial density of the air, and $w (t)$ is the ionization
rate {\cite{ammosov_tunnel_1986}}. Thus,
$\ensuremath{\boldsymbol{\mathcal{E}}}_{\ensuremath{\operatorname{THz}}} (t)
\propto N_g  \left( 1 - \exp \left[ - \int^t \mathrm{d} t' w (t') \right]
\right) \ensuremath{\boldsymbol{E}} (t)$ accounts for the residual current
induced by the external field in the plasma. The expansion up to the first
order of the exponent, i.e.,
\begin{equation}
  \ensuremath{\boldsymbol{\mathcal{E}}}_{\ensuremath{\operatorname{THz}}} (t)
  \propto \ensuremath{\boldsymbol{E}} (t)  \int^t \mathrm{d} t' w (t'),
  \label{eq:PC-Et}
\end{equation}
however, is essentially based on the single-atomic ionization, since Eq.
(\ref{eq:PC-Et}) is the solution of $\partial_t N (t) = N_g w (t)$, where the
depletion of the neutral atoms in plasma, $- N (t) w (t)$, as appeared in Eq.
(\ref{eq:PC-Nt}), is neglected. Therefore, Eq. (\ref{eq:PC-Et}) is referred to
as the {\itshape{single-atom}} PC (SPC) model (see Supplementary S5 for
comparison of results between PC and SPC). In Fig. \ref{fig:pp-comparison}(c),
the distribution from SPC model also shows a good agreement with the
experiment in Fig. \ref{fig:pp-comparison}(a).

\begin{figure}[h]
  \includegraphics[scale=1]{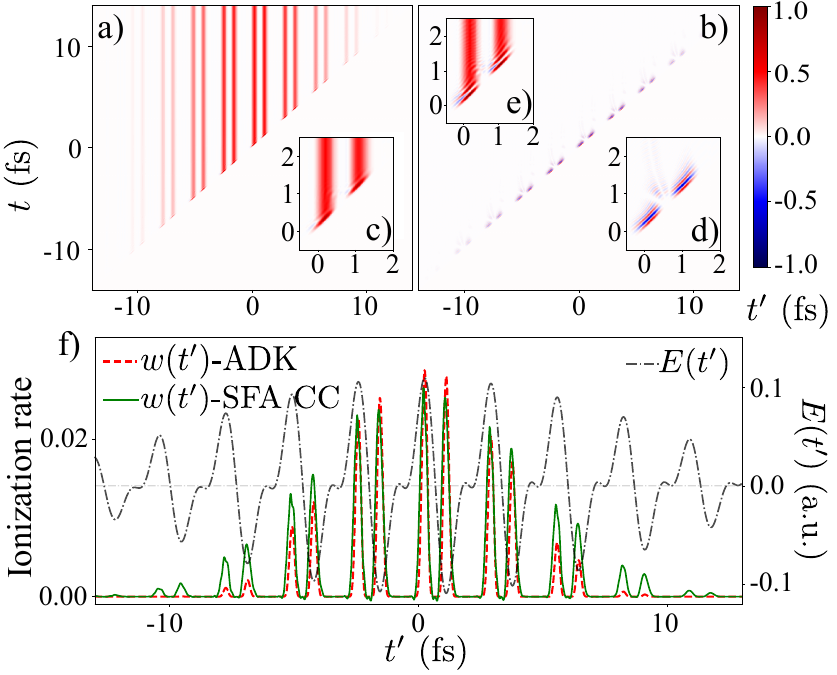}
  \caption{The $w (t' ; t)$ of the SFA-CC and the comparison with $w (t')$ of
  the SPC when $\theta = 0${\textdegree} and $\tau = 0.33 \text{ fs}$. In
  collinear dual-color laser fields, the SFA-CC derived $w_1 (t' ; t)$ and
  $w_2 (t' ; t)$ are shown in (a) and (b), respectively, with their detailed
  zoom-in around $t' = 0$ in insets (c) and (d). The total contribution, $w
  (t' ; t) = w_1 (t' ; t) + w_2 (t' ; t)$, is presented in inset (e), showing
  that $w_2 (t' ; t)$ almost contributes at $t' = t$ only. In (f), $w (t' ; t
  \rightarrow \infty)$ of the SFA-CC (solid line), is compared with $w (t')$
  of the SPC (dashed line), showing the correspondence between the SFA based
  quantum model and the semi-classical PC model.\label{fig:eq-comparison}}
\end{figure}

It is not a coincidence that all these models agree well with the experimental
results. As is shown in the followings, there exists a linkage among different
theories. Obviously, the SFA-CC and the SPC share the similar form---the rate
$w (t')$ in Eq. (\ref{eq:PC-Et}) is simply substituted by $w_1 (t' ; t)$ in
Eq. (\ref{eq:a1}), as defined by $w_1 (t' ; t) = \int_{t_i}^t \mathrm{d} t''
\eta (t', t'') W (t') W^{\ast} (t'') \mathrm{e}^{\mathrm{i}
S_{\ensuremath{\boldsymbol{k}}', I_{\text{p}}} (t', t'')}$. Before showing the
correspondence between $w (t)$ and $w_1 (t' ; t)$, we first examine the
distribution of $w_1 (t' ; t)$ versus the ionization time $t'$ at different
emission instants $t$. As is presented in Fig. \ref{fig:eq-comparison}(a) and
inset (c), $w_1 (t' ; t)$ is nonvanishing only for $t > t'$, as is restricted
by the principle of causality that the ionization event should precede the
emission. Despite the apparent dependence of $w_1 (t' ; t)$ on $t$, the
temporal distributions for $t' < t$ remain almost unaltered versus $t$, except
around the boundary when $t = t'$.

Besides $w_1 (t' ; t)$ for $\ensuremath{\boldsymbol{a}}_1$, we can also define
$w_2 (t' ; t)$ from $\ensuremath{\boldsymbol{a}}_2$ so that the total
contribution of the SFA-CC, $w (t' ; t) = w_1 (t' ; t) + w_2 (t' ; t)$, is
formally consistent with $w (t)$ in Eq. (\ref{eq:PC-Et}). For the collinear
dual-color laser fields, straightforwardly we have $w_2 (t' ; t) = - 2
\text{Re} \eta (t, t') W (t) W^{\ast} (t')  \frac{k' + A (t)}{E (t)} 
\mathrm{e}^{\mathrm{i} S_{\ensuremath{\boldsymbol{k}}', I_p} (t, t')}$. The
distribution of $w_2 (t' ; t)$ is shown in Fig. \ref{fig:eq-comparison}(b) and
inset (d). The contribution from $w_2 (t' ; t)$ is almost negligible, except
when $t' \rightarrow t$, that is why it is referred to as the boundary term.
It may influence some details of the TWG process, leading to subtle
differences between the quantum mechanical SFA-CC and the semiclassical PC
models. When $t$ is sufficiently large, the distribution of $w (t' ; t)$ is
almost equivalent to $w_1 (t' ; t)$.

With the approximation that the contribution from $w_2 (t' ; t)$ negligible
and $w_1 (t' ; t)$ roughly independent of $t$, the $w_1 (t' ; t)$ versus the
ionization time $t'$ can be directly compared with $w (t')$, as shown in Fig.
\ref{fig:eq-comparison}(f). Comparing SFA-CC with SPC model, if the emission
time $t$ is sufficiently away from $t'$, $w_1 (t' ; t)$ versus $t'$ presents
the similar distribution as $w (t')$ of the SPC. In other words, $w (t')$ can
be considered as the quasi-static limit of the $w_1 (t' ; t)$ restricted by
the time ordering $t' < t$, even though $w (t')$ is introduced from the view
of macroscopic photoelectric current, while $w_1 (t' ; t)$ is derived
completely from single-atom based microscopic process of strong field
ionization. In explaining the TWG, the SFA and (S)PC theories exhibit the
quantum-classical correspondence.

Besides the formal similarities to the PC model,
$\ensuremath{\boldsymbol{a}}_1 (t)$ in Eq. (\ref{eq:a1}) explicitly shows the
third-order dependence on the external electric field
$\ensuremath{\boldsymbol{E}} (t)$, as presented by the perturbative
third-order response in the FWM model. The response to the incident fields, as
predicted in the FWM, can be verified by experimental observations of
polarization- and intensity-dependence THz yields
{\cite{Zhang2009OL,Xie2006PRL}}. The conventional perturbative FWM
susceptibility, however, is replaced by the nonperturbative transition dipole
moment induced by strong fields. Thus, the SFA-CC is in concordance with the
ionization induced multiwave mixing {\cite{Kostinprl2016}}, unifying the
existent explanations including FWM the PC models.

In conclusion, the mechanism of the dual-color TWG has been clarified under
the theoretical framework of the strong field physics, claiming another
success of the renowned SFA theory. Although the ensemble behavior of the
laser-induced plasma may influence the radiation yields, the underlying origin
of the TWG resides within the scope of nonperturbative single atomic strong
field processes. In contrast to the HHG emitted by the CB transition of a
recolliding electron, the TWG originates from the CC transition of a released
electron after the ionization. The TWG mechanism of the CC---the "soft"
transition---beyond the HHG mechanism of "hard" recollision, is a complement
to the radiation theory of the strong field physics. Meanwhile, it is shown
that the classical PC model can be derived from the SFA-CC method, bridging
between the classical and quantum-mechanical interpretations. Also, the FWM
can be reached from the SFA-CC by presenting the explicit third-order
dependence on the electric field. Hence, the SFA-CC serves to unify the FWM
and PC models, while it offers more microscopic details comparing to the
latter coarse-grained models. Our research of the TWG mechanism opens up the
possibility to extract the ultrafast dynamics of continuum electron from the
ionization-induced THz emission.

The study was supported by National Natural Science Foundation of China (NSFC)
(11420101003, 11604347, 11827806, 11874368, 61675213, 91636105). We also
acknowledge the support from Shanghai-XFEL beamline project (SBP) and Shanghai
HIgh repetitioN rate XFEL and Extreme light facility (SHINE).

\bibliographystyle{apsrev4-1}
\bibliography{main}

\pagebreak
\widetext
\begin{center}
\textbf{\large Supplemental Materials: Dual-Color Laser Induced Terahertz Generation in Strong Field Approximation}
\end{center}
\setcounter{section}{0}
\setcounter{equation}{0}
\setcounter{figure}{0}
\setcounter{table}{0}
\setcounter{page}{1}
\makeatletter
\renewcommand{\thesection}{S\arabic{section}}
\renewcommand{\theequation}{S\arabic{equation}}
\renewcommand{\thefigure}{S\arabic{figure}}
\renewcommand{\bibnumfmt}[1]{[S#1]}
\renewcommand{\citenumfont}[1]{S#1}

\section{Transition Dipole Moment under the Strong Field Approximation}

The expected value of dipole moment is $\ensuremath{\boldsymbol{d}} (t) = -
\langle \Psi (t) | \hat{\ensuremath{\boldsymbol{r}}} | \Psi (t) \rangle = -
\langle \Psi_0 (t_i) | \hat{U} (t_i, t)  \hat{\ensuremath{\boldsymbol{r}}} 
\hat{U} (t, t_i) | \Psi_0 (t_i) \rangle$. With the time evolution operator
$\hat{U} (t, t_i)$ expanded by Dyson series, $\hat{U} (t, t_i) = \hat{U}_0 (t,
t_i) - \mathrm{i} \int_{t_i}^t \mathrm{d} t'  \hat{U} (t, t')  \hat{W} (t') 
\hat{U}_0 (t', t_i)$, where $\hat{U}_0 (t', t_i)$ is the interaction-free time
evolution operator and $\hat{W}$ is the interaction operator, we find that
$\ensuremath{\boldsymbol{d}} (t) =\ensuremath{\boldsymbol{d}}^{(0)} (t)
+\ensuremath{\boldsymbol{d}}^{(1)} (t) +\ensuremath{\boldsymbol{d}}^{(2)}
(t)$, where
\begin{eqnarray}
  \ensuremath{\boldsymbol{d}}^{(0)} (t) & = & - \langle \Psi_0 (t) |
  \hat{\ensuremath{\boldsymbol{r}}} | \Psi_0 (t) \rangle,  \label{eq:d0}\\
  \ensuremath{\boldsymbol{d}}^{(1)} (t) & = & - (- \mathrm{i}) \int_{t_i}^t
  \mathrm{d} t' \langle \Psi_0 (t) | \hat{\ensuremath{\boldsymbol{r}}} 
  \hat{U} (t, t')  \hat{W} (t') | \Psi_0 (t') \rangle + \text{c.c.}, 
  \label{eq:d1}\\
  \ensuremath{\boldsymbol{d}}^{(2)} (t) & = & - \int_{t_i}^t \mathrm{d} t'' 
  \int_{t_i}^t \mathrm{d} t' \langle \Psi_0 (t'') | \hat{W} (t'')  \hat{U}
  (t'', t)  \hat{\ensuremath{\boldsymbol{r}}}  \hat{U} (t, t')  \hat{W} (t') |
  \Psi_0 (t') \rangle .  \label{eq:d2}
\end{eqnarray}
The $\ensuremath{\boldsymbol{d}}^{(0)} (t)$ vanishes in a spherically
symmetric system. The $\ensuremath{\boldsymbol{d}}^{(1)} (t)$, referred to as
the continuum-bound (CB) transition dipole moment, depicts the coherent
emission at time $t$ induced by the transition of electron from continuum,
which accumulates over all possible ionization events at time $t'$, to the
bound state. The $\ensuremath{\boldsymbol{d}}^{(1)} (t)$ is widely recognized
to dominate the high-order harmonic generation (HHG) process. Similarly, the
$\ensuremath{\boldsymbol{d}}^{(2)} (t)$, referred to as the
continuum-continuum (CC) transition dipole moment, describes the coherent
emission induced by the transition between states of continuum, which is often
negligible for the HHG calculation. Here, however, we emphasize the role of
the $\ensuremath{\boldsymbol{d}}^{(2)} (t)$ in terahertz wave generation
(TWG). In the followings, the details of $\ensuremath{\boldsymbol{d}}^{(1)}
(t)$ and $\ensuremath{\boldsymbol{d}}^{(2)} (t)$ as used to evaluate the
emission are presented.

\subsection{$d^{(1)}$: Continuum-Bound (CB) Transition}\label{sec:d1}

The CB transition $\ensuremath{\boldsymbol{d}}^{(1)} (t)$ is given by Eq.
(\ref{eq:d1}). Under the strong field approximation (SFA) which neglects the
interaction between the photoelectron and the parent ion, the full
time-evolution operator is substituted by the operator with the external light
field only, $\hat{U} (t, t') \rightarrow \hat{U}_I (t, t')$, yielding
\[ \ensuremath{\boldsymbol{d}}^{(1)} (t) = - (- \mathrm{i}) \int_{t_i}^t
   \mathrm{d} t' \langle \Psi_0 (t) | \hat{\ensuremath{\boldsymbol{r}}} 
   \hat{U}_I (t, t')  \hat{W} (t') | \Psi_0 (t') \rangle + \text{c.c.} . \]
with $\hat{U}_I (t, t') = \int \mathrm{d} \ensuremath{\boldsymbol{k}}|
\Psi^{(V)}_{\ensuremath{\boldsymbol{k}}} (t) \rangle \langle
\Psi^{(V)}_{\ensuremath{\boldsymbol{k}}} (t') |$. The Volkov state $|
\Psi^{(V)}_{\ensuremath{\boldsymbol{k}}} (t) \rangle =
|\ensuremath{\boldsymbol{k}}+\ensuremath{\boldsymbol{A}} (t) \rangle
\mathrm{e}^{- \mathrm{i} S_{\ensuremath{\boldsymbol{k}}} (t)}$ with
$S_{\ensuremath{\boldsymbol{k}}} (t) = \frac{1}{2} \int^t \mathrm{d} t' 
[\ensuremath{\boldsymbol{k}}+\ensuremath{\boldsymbol{A}} (t')]^2$ and the
vector potential $\ensuremath{\boldsymbol{A}} (t)$. Considereing the
interaction $W (t') =\ensuremath{\boldsymbol{\mu}}
[\ensuremath{\boldsymbol{k}}' +\ensuremath{\boldsymbol{A}}(t')] \cdot
\ensuremath{\boldsymbol{E}} (t')$ between the electron with the incident
electric field $\ensuremath{\boldsymbol{E}} (t')$ and substituting $| \Psi_0
(t') \rangle = | \psi_0 \rangle \mathrm{e}^{- \mathrm{i} E_0 t'}$, it is shown
that $\ensuremath{\boldsymbol{d}}^{(1)} (t) = \mathrm{i} \int_{t_i}^t
\mathrm{d} t'  \int \mathrm{d} \ensuremath{\boldsymbol{k}} \mathrm{e}^{-
\mathrm{i} S_{\ensuremath{\boldsymbol{k}}, I_{\text{p}}} (t, t')}
\ensuremath{\boldsymbol{\mu}}^{\ast} 
[\ensuremath{\boldsymbol{k}}+\ensuremath{\boldsymbol{A}}(t)]
\ensuremath{\boldsymbol{E}}^{\ast} (t') \cdot \ensuremath{\boldsymbol{\mu}}
[\ensuremath{\boldsymbol{k}}+\ensuremath{\boldsymbol{A}}(t')] + \text{c.c.}$,
where $\ensuremath{\boldsymbol{k}}$ is the intermediate momentum, and
$S_{\ensuremath{\boldsymbol{k}}, I_{\text{p}}} (t, t') = \int_{t'}^t
\mathrm{d} t''  (\frac{1}{2}
[\ensuremath{\boldsymbol{k}}+\ensuremath{\boldsymbol{A}}(t'')]^2 +
I_{\text{p}})$ with the ionization energy $I_{\text{p}} = - E_0$. In this
work, the 1s state of the hydrogen atom is considered as the initial state for
simplicity, the dipole matrix element reads $\ensuremath{\boldsymbol{\mu}}
(\ensuremath{\boldsymbol{k}}) = - \mathrm{i} 2^{\frac{7}{2}}
\ensuremath{\boldsymbol{k}}/ [\pi (k^2 + 1)^3]$, and $I_{\text{p}} = 0.5$
a.u.. The integration over momentum can be approximated with the stationary
phase, and the stationary point $\ensuremath{\boldsymbol{k}}_s$ is the
solution to the equation $\nabla_{\ensuremath{\boldsymbol{k}}}
S_{\ensuremath{\boldsymbol{k}}, I_{\text{p}}} (t, t') \overset{!}{=} 0$.
Therefore, we obtain the CB transition dipole
\begin{equation}
  \ensuremath{\boldsymbol{d}}^{(1)} (t) = \mathrm{i} \int_{t_i}^t dt' (\frac{2
  \pi}{\mathrm{i} (t - t')})^{\frac{3}{2}}
  \ensuremath{\boldsymbol{\mu}}^{\ast}  [\ensuremath{\boldsymbol{k}}_s
  +\ensuremath{\boldsymbol{A}}(t)] \ensuremath{\boldsymbol{E}}^{\ast} (t')
  \cdot \ensuremath{\boldsymbol{\mu}} [\ensuremath{\boldsymbol{k}}_s
  +\ensuremath{\boldsymbol{A}}(t')] \mathrm{e}^{- \mathrm{i}
  S_{\ensuremath{\boldsymbol{k}}_s, I_{\text{p}}} (t, t')} + \text{c.c.} .
  \label{eq:d1-result}
\end{equation}
\subsection{$d^{(2)}$: Continuum-Continuum (CC) Transition }\label{sec:d2}

Under the strong field approximation, the substitution $\hat{U} (t, t')
\rightarrow \hat{U}_I (t, t')$ yields
\begin{eqnarray}
  \ensuremath{\boldsymbol{d}}^{(2)} (t) & \simeq & - (+ \mathrm{i}) (-
  \mathrm{i}) \int_{t_i}^t \mathrm{d} t''  \int_{t_i}^t \mathrm{d} t' \langle
  \Psi_0 (t'') | \hat{W} (t'')  \hat{U}_I (t'', t) 
  \hat{\ensuremath{\boldsymbol{r}}}  \hat{U}_I (t, t')  \hat{W} (t') | \Psi_0
  (t') \rangle . 
\end{eqnarray}
Similar to the manipulation for the CB transition, the expansion arrives
\begin{eqnarray}
  \ensuremath{\boldsymbol{d}}^{(2)} (t) & = & - (+ \mathrm{i}) (- \mathrm{i})
  \int_{t_i}^t \mathrm{d} t''  \int_{t_i}^t \mathrm{d} t'  \int \mathrm{d}
  \ensuremath{\boldsymbol{k}}'  \int \mathrm{d} \ensuremath{\boldsymbol{k}}''
  \nonumber\\
  &  & \times \ensuremath{\boldsymbol{\mu}}^{\ast}
  [\ensuremath{\boldsymbol{k}}'' +\ensuremath{\boldsymbol{A}} (t'')] \cdot
  \ensuremath{\boldsymbol{E}} (t'') \langle \ensuremath{\boldsymbol{k}}''
  +\ensuremath{\boldsymbol{A}} (t) | \hat{\ensuremath{\boldsymbol{r}}}
  |\ensuremath{\boldsymbol{k}}' +\ensuremath{\boldsymbol{A}} (t) \rangle
  \nonumber\\
  &  & \times \ensuremath{\boldsymbol{\mu}} [ \ensuremath{\boldsymbol{k}}'
  +\ensuremath{\boldsymbol{A}}  (t')] \cdot \ensuremath{\boldsymbol{E}} (t')
  \mathrm{e}^{- \mathrm{i} S_{\ensuremath{\boldsymbol{k}}'', I_{\text{p}}}
  (t'', t)} \mathrm{e}^{\mathrm{i} S_{\ensuremath{\boldsymbol{k}}',
  I_{\text{p}}} (t', t)} . 
\end{eqnarray}
After applying $\langle \ensuremath{\boldsymbol{k}}''
+\ensuremath{\boldsymbol{A}} (t) | \hat{\ensuremath{\boldsymbol{r}}}
|\ensuremath{\boldsymbol{k}}' +\ensuremath{\boldsymbol{A}} (t) \rangle =
\mathrm{i} \nabla_{\ensuremath{\boldsymbol{k}}''} \delta
(\ensuremath{\boldsymbol{k}}'' -\ensuremath{\boldsymbol{k}}')$, the further
derivation shows
\begin{eqnarray}
  \ensuremath{\boldsymbol{d}}^{(2)} (t) & = & - (+ \mathrm{i}) (- \mathrm{i})
  (\mathrm{i}) (- 1) \int_{t_i}^t \mathrm{d} t''  \int_{t_i}^t \mathrm{d} t' 
  \int \mathrm{d} \ensuremath{\boldsymbol{k}}' \ensuremath{\boldsymbol{\mu}}
  [\ensuremath{\boldsymbol{k}}' +\ensuremath{\boldsymbol{A}} (t')] \cdot
  \ensuremath{\boldsymbol{E}} (t') \nonumber\\
  &  & \times \left[ \nabla_{\ensuremath{\boldsymbol{k}}'} - \mathrm{i}
  \nabla_{\ensuremath{\boldsymbol{k}}'} S_{\ensuremath{\boldsymbol{k}}',
  I_{\text{p}}} (t'', t) \right] \{ \ensuremath{\boldsymbol{\mu}}^{\ast}
  [\ensuremath{\boldsymbol{k}}' +\ensuremath{\boldsymbol{A}} (t'')] \cdot
  \ensuremath{\boldsymbol{E}} (t'') \} \mathrm{e}^{\mathrm{i}
  S_{\ensuremath{\boldsymbol{k}}', I_{\text{p}}} (t', t'')} . 
\end{eqnarray}
The integration over $\ensuremath{\boldsymbol{k}}'$ can also be treated by the
stationary phase approximation. Solving the saddle point equation
$\nabla_{\ensuremath{\boldsymbol{k}}'} S_{\ensuremath{\boldsymbol{k}}',
I_{\text{p}}} (t', t'') = 0$, we obtain $\ensuremath{\boldsymbol{k}}'_s \equiv
\ensuremath{\boldsymbol{k}}'_s (t', t'') = - [\ensuremath{\boldsymbol{\alpha}}
(t') -\ensuremath{\boldsymbol{\alpha}} (t'')] / (t' - t'')$ with the excursion
$\ensuremath{\boldsymbol{\alpha}} (t) = \int^t \mathrm{d} t'
\ensuremath{\boldsymbol{A}} (t')$. The approximation with
$\ensuremath{\boldsymbol{k}}_s'$ results in
\begin{eqnarray}
  \ensuremath{\boldsymbol{d}}^{(2)} (t) & = & - (+ \mathrm{i}) (- \mathrm{i})
  (\mathrm{i}) (- 1) \int_{t_i}^t \mathrm{d} t''  \int_{t_i}^t \mathrm{d} t' 
  \left( \frac{2 \pi}{\mathrm{i} (t' - t'')} \right)^{3 / 2}
  \ensuremath{\boldsymbol{\mu}} [\ensuremath{\boldsymbol{k}}'_s
  +\ensuremath{\boldsymbol{A}} (t')] \cdot \ensuremath{\boldsymbol{E}} (t') 
  \nonumber\\
  &  & \times \left\{ \nabla_{\ensuremath{\boldsymbol{k}}'_s}
  \ensuremath{\boldsymbol{\mu}}^{\ast} [\ensuremath{\boldsymbol{k}}'_s
  +\ensuremath{\boldsymbol{A}} (t'')] - \mathrm{i}
  \ensuremath{\boldsymbol{\mu}}^{\ast} [\ensuremath{\boldsymbol{k}}'_s
  +\ensuremath{\boldsymbol{A}} (t'')] \nabla_{\ensuremath{\boldsymbol{k}}'_s}
  S_{\ensuremath{\boldsymbol{k}}'_s, I_{\text{p}}} (t'', t)  \right\} \cdot
  \ensuremath{\boldsymbol{E}} (t'') \mathrm{e}^{\mathrm{i}
  S_{\ensuremath{\boldsymbol{k}}'_s, I_{\text{p}}} (t', t'')} . 
\end{eqnarray}
Within the curly bracket, the first term of
$\nabla_{\ensuremath{\boldsymbol{k}}_s'} \ensuremath{\boldsymbol{\mu}}^{\ast}$
is negligible. Substituting $\nabla_{\ensuremath{\boldsymbol{k}}'_s}
S_{\ensuremath{\boldsymbol{k}}'_s, I_{\text{p}}} (t'', t)
=\ensuremath{\boldsymbol{k}}'_s (t'' - t) +\ensuremath{\boldsymbol{\alpha}}
(t'') -\ensuremath{\boldsymbol{\alpha}} (t)$ in the second term,
\begin{eqnarray}
  \ensuremath{\boldsymbol{d}}^{(2)} (t) & \simeq & \int_{t_i}^t \mathrm{d} t''
  \int_{t_i}^t \mathrm{d} t'  \left( \frac{2 \pi}{\mathrm{i} (t' - t'')}
  \right)^{3 / 2}  \nonumber\\
  &  & \ensuremath{\boldsymbol{\mu}} [\ensuremath{\boldsymbol{k}}'_s
  +\ensuremath{\boldsymbol{A}} (t')] \cdot \ensuremath{\boldsymbol{E}} (t')
  \ensuremath{\boldsymbol{\mu}}^{\ast} [\ensuremath{\boldsymbol{k}}'_s
  +\ensuremath{\boldsymbol{A}} (t'')] \cdot \ensuremath{\boldsymbol{E}} (t'')
  \nonumber\\
  &  & \times \{ \ensuremath{\boldsymbol{k}}'_s (t'' - t)
  +\ensuremath{\boldsymbol{\alpha}} (t'') -\ensuremath{\boldsymbol{\alpha}}
  (t) \} \mathrm{e}^{\mathrm{i} S_{\ensuremath{\boldsymbol{k}}'_s,
  I_{\text{p}}} (t', t'')} . 
\end{eqnarray}
The emission is given by the acceleration $\ensuremath{\boldsymbol{a}}=
\ddot{\ensuremath{\boldsymbol{d}}}^{(2)} (t)$. Applying the Leibniz integral
rule, we evaluate the second-order derivative of the dipole moment with
respect to $t$,
\begin{eqnarray}
  \ddot{\ensuremath{\boldsymbol{d}}}^{(2)} (t) & = &
  \ensuremath{\boldsymbol{E}} (t) \int_{t_i}^t \mathrm{d} t'' \int_{t_i}^t
  \mathrm{d} t'  \left( \frac{2 \pi}{\mathrm{i} (t' - t'')} \right)^{3 / 2}
  \mathrm{e}^{\mathrm{i} S_{\ensuremath{\boldsymbol{k}}'_s (t', t''), I_p}
  (t', t'')} \nonumber\\
  &  & \times \ensuremath{\boldsymbol{\mu}} [\ensuremath{\boldsymbol{k}}'_s
  (t', t'') +\ensuremath{\boldsymbol{A}} (t')] \cdot
  \ensuremath{\boldsymbol{E}} (t') \ensuremath{\boldsymbol{\mu}}^{\ast}
  [\ensuremath{\boldsymbol{k}}'_s (t', t'') +\ensuremath{\boldsymbol{A}}
  (t'')] \cdot \ensuremath{\boldsymbol{E}} (t'') \nonumber\\
  &  & - 2 \text{Re} \int_{t_i}^t \mathrm{d} t'  \left( \frac{2
  \pi}{\mathrm{i} (t - t')} \right)^{3 / 2} \mathrm{e}^{\mathrm{i}
  S_{\ensuremath{\boldsymbol{k}}_s' (t, t'), I_p} (t, t')}
  \ensuremath{\boldsymbol{\mu}} [\ensuremath{\boldsymbol{k}}'_s (t, t')
  +\ensuremath{\boldsymbol{A}} (t)] \cdot \ensuremath{\boldsymbol{E}} (t)
  \nonumber\\
  &  & \times \ensuremath{\boldsymbol{\mu}}^{\ast}
  [\ensuremath{\boldsymbol{k}}'_s (t, t') +\ensuremath{\boldsymbol{A}} (t')]
  \cdot \ensuremath{\boldsymbol{E}} (t')  [\ensuremath{\boldsymbol{k}}'_s (t,
  t') +\ensuremath{\boldsymbol{A}} (t)], 
\end{eqnarray}
which is exactly the full form of $\ddot{\ensuremath{\boldsymbol{d}}}^{(2)}
(t) =\ensuremath{\boldsymbol{a}}_1 (t) +\ensuremath{\boldsymbol{a}}_2 (t)$ as
presented by Eqs. (2) and (3) in the main text.

\section{Experiment}

The femtosecond amplifier (Libra, Coherent Inc.) delivers a laser pulse with
$\sim$800 nm center wavelength and $\sim$35 fs pulse duration. The pulse with
$\sim$1.75 mJ is guided into the experiment setup (Fig. S1). The beam with
96\% pulse energy is reflected as the pump beam for the terahertz wave
generation (TWG), and the transmission beam is used as the probe beam for
electro-optic sampling (EOS). The pump beam passes through a 200-$\mathrm{\mu
m}$ type-I BBO crystal with the double-frequency efficiency of $\sim$23\%. The
polarization of the laser beam is horizontal ($p$-polarized),
and{\color{black}{ the optical axis of BBO is kept perpendicular with the
laser polarization }}to obtain the maximum efficiency. The outcoming second
harmonic beam is $s$-polarized. The relative polarization $\theta$ between the
fundamental $\omega$ and second harmonic $2 \omega$ beams can be controlled by
rotating the zero-order dual-wavelength wave plate (DWP), which acts as a
half-wave plate for the $\omega$ beam and a full-wave plate for the $2 \omega$
beam. The defination of observables is schematically illustrated in Fig. S2.
Controlling $\theta$ with a half-wave plate, instead of rotating BBO crystal,
avoids the mixture of the polarization of $\ensuremath{\boldsymbol{{\^o}}}$
ray and $\ensuremath{\boldsymbol{{\^e}}}$ ray in the BBO crystal.
{\color{black}{The ellipticities}} of both $2 \omega$ and $\omega$ beams are
better than 0.1 when DWP rotating. Throughout the measurement, the $\omega$
and $2 \omega$ beams can be considered as linear polarized. The time delay
$\tau$ of the dual-color fields can be varied by moving the BBO crystal along
the propagation direction, because of different refractive indices at $\omega$
and 2$\omega$ in air. Because of the collinear propagation geometry, the
fluctuation of $\tau$ can be passively suppressed within the sub-wavelength
accuracy.

The dual-color fields are focused by a silver parabolic mirror with an
effective focal length of $\sim$100 mm. Atmospheric air is ionized for the
TWG, simultaneously emitting third-order harmonic generation (THG). Here, we
use a tightly-focusing scheme to delibrately prevent the propagation effect in
plasma. The THz waves are collected and collimated with a gold parabolic
mirror with $\sim$100-mm focal length, and focused into 1-mm-thick (110)-cut
ZnTe crystal with a same parabolic mirror. A 500-$\mu m$-thick polished
silicon wafer reflects the residual laser, allowing the transmission of only
THz component. A pellicle beam splitter combines the THz pulse and the probe
beam to implement the free-space EOS detection. The signal-to-noise ratio
(SNR) of the terahertz time-domain waveforms is better than 100:1. A
polarization-sensitive THz-EOS is employed in our experiment. A metal
wire-grid THz polarizer filters out the orthogonally-polarized components of
TWG, and the ZnTe crystal is fixed at the special orientation, where the
responses for $s$- and $p$-polarized THz components are the same.

In the measurement, both $s$- and $p$-polarized THz electric fields
$\mathcal{E}_{\text{THz}} (t)$ are recorded with the EOS method. The THz
peak-peak (PP) amplitude is defined as $S_{\ensuremath{\operatorname{THz}}} =
\pm \mid \max [\mathcal{E}_{\text{THz}} (t)] - \min [\mathcal{E}_{\text{THz}}
(t)] \mid$. The distributions of $S_{\ensuremath{\operatorname{THz}}, s (p)}
(\tau, \theta)$ are presented in Fig. 1(a) in the main text. The
$S_{\ensuremath{\operatorname{THz}}, s (p)} (\tau, \theta)$ are normalized to
the maximum. The positive direction of $S_{\ensuremath{\operatorname{THz}}, s}
(\tau, \theta)$ and $S_{\ensuremath{\operatorname{THz}}, p} (\tau, \theta)$
are defined when the maxima of THz waveforms appear along the positive
direction of $s$ and $p$ axes.

The THG of $\sim$266 nm is coincidently measured with the TWG for the
{\itshape{in situ}} determination of the absolute $\tau$ of the dual-color
fields. The THG signals are reflected by the silicon wafer with residual
$\omega$ and $2 \omega$ beams, and spectrally separated by a suprasil prism.
The $s$- and $p$-components of THG signals are decomposed with a glan-laser
polarizer, and collected into a fiber spectrometer. The THG signals are
measured with 50-ms integration time, 10-time average in our measurement.

\begin{figure}[h]
  \includegraphics{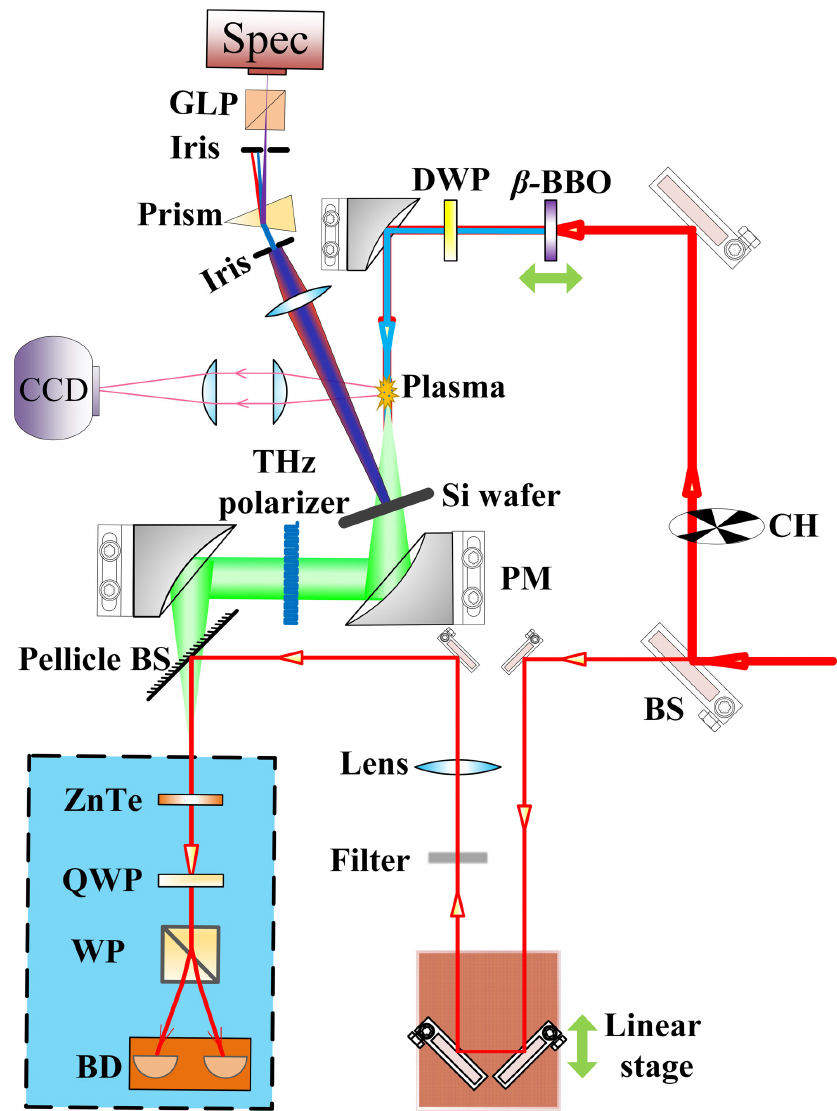}
  \caption{Experimental setup. BS: beam splitter; CH: chopper; $\beta$-BBO:
  beta barium borate; DWP: dual-wavelength plate; PM: parabolic mirror; QWP:
  quarter wave plate; GLP: glan-laser polarizer; WP: wollaston polarizer; BD:
  balanced detector.}
\end{figure}

\begin{figure}[h]
  \includegraphics{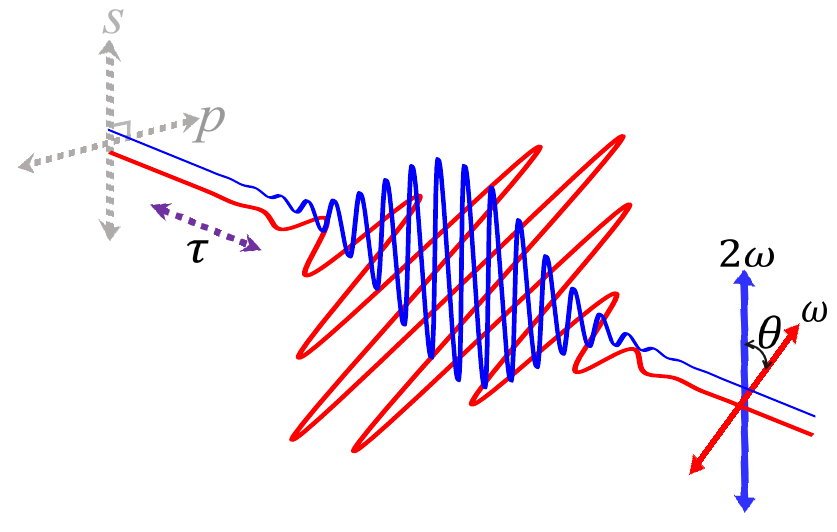}
  \caption{Definition of observables. The 800 nm ($\omega$) and 400 nm ($2
  \omega$) beams collinearly propagate. The polarization of the $2 \omega$
  beam is always $s$-polarized. The relative time delay $\tau$ and
  polarization $\theta$ are controlled in the measurement. Both the $s$-and
  $p$-polarized terahertz waves are detected.}
\end{figure}

\section{Determination of Time Delay Zero by Joint Measurement of Third-Order
Harmonics}

The dependence of TWG on time delay $\tau$ between the dual-color fields is
critical to determine the TWG mechanisms. For instance, the TWG maximum is
predicted to appear at $\tau = 0.33$ fs in the photocurrent (PC) theory,
however, at $\tau = 0$ fs in the perturbative four-wave-mixing (FWM) theory.
Only when the $\tau$ is precisely known, the electric field waveforms for the
TWG can be determined for further comparison of the measured data with
different theories. However, the precise $\tau$ is difficult to obtain, since
it is nontrivial to direct monitor the electric fields in practical
experiments.

In our experiment, the joint measurement of TWG and THG are conducted. The THG
yields along $s$-polarization $I_{\text{3rd}} (\tau, \theta)$ are shown in
Fig. S3(a). In addition, we have examined the $I_{\text{3rd}} (\tau, \theta)$
evaluated by different theories, including the CB, CC transitions and SPC. All
results, as shown in Fig. S3(b)-(d), predict the similar $\tau$ dependence
that the maximum $I_{\text{3rd}} (\tau)$ appears at $\tau = 0$ fs. The time
delay zero of dual-color fields in the experiment can therefore be precisely
determined by comparison with the theoretical results.

\begin{figure}[h]
  \includegraphics{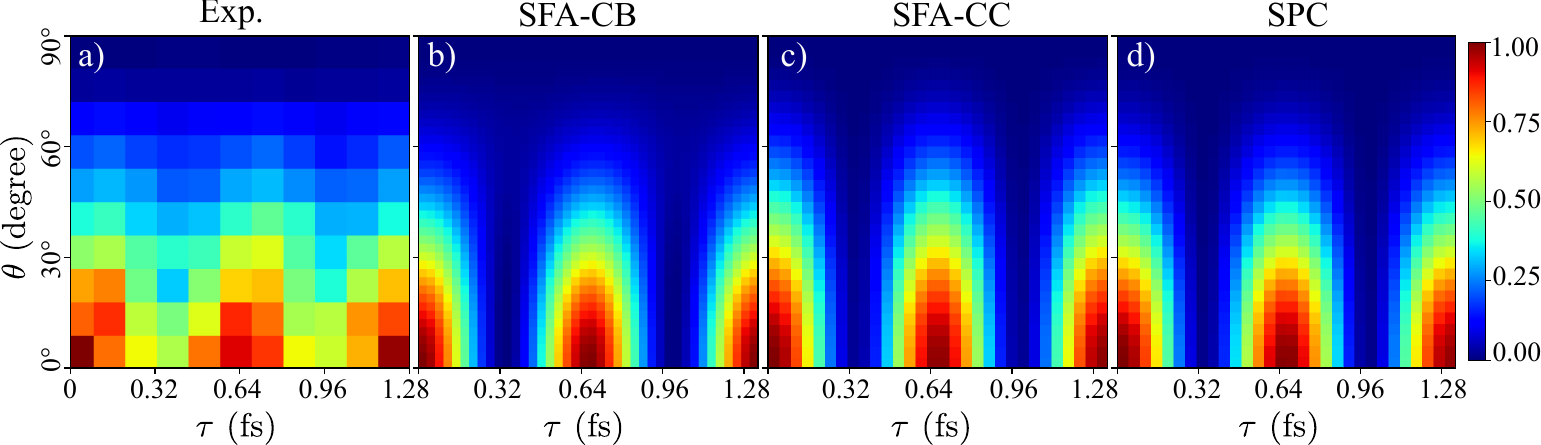}
  \caption{The distribution of the THG along $s$-polarization $I_{\text{3rd}}
  (\tau, \theta)$ obtained from (a) the measurement, (b)
  $\ddot{\ensuremath{\boldsymbol{d}}}^{(1)} (t)$ of \ SFA-CB , (c)
  $\ddot{\ensuremath{\boldsymbol{d}}}^{(2)} (t)$ of SFA-CC, and (d) SPC.}
\end{figure}

\section{Continuum-Bound Transition in Strong Field Approximation}

The CB transition is evaluated by Eq. (\ref{eq:d1-result}) using the same
parameters as in the SFA-CC calculation. For the dual-color laser fields with
time delay $\tau$, the electric field $\ensuremath{\boldsymbol{E}} (t)$ is
given by $\ensuremath{\boldsymbol{E}} (t)
=\ensuremath{\boldsymbol{E}}_{\omega} (t) +\ensuremath{\boldsymbol{E}}_{2
\omega}  (t - \tau)$. The $\sin^2$-envelope is used for the construction of
femtosecond pulses. The fundamental field includes 24 cycles with the strength
of 0.08 a.u.. The second-harmonic field includes 48 cycles with the strength
of 0.046 a.u., assuming BBO has the conversion efficiency $\sim$30\%. The
$\ddot{\ensuremath{\boldsymbol{d}}}^{(1)} (t)$ is Fourier transformed into the
frequency domain, and the THz and the third-order hamonic components are
filtered out.

The parameteric dependence $S_{\ensuremath{\operatorname{THz}}, s (p)} (\tau,
\theta)$ predicted by CB transition is shown in Fig. S4.

\begin{figure}[h]
  \includegraphics{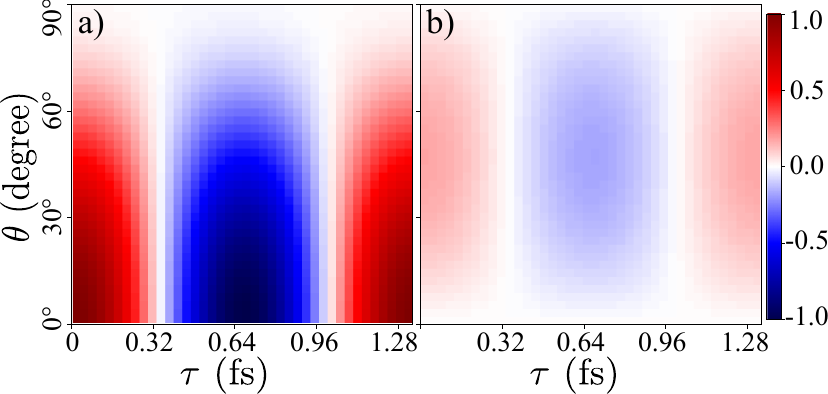}
  \caption{The prediction of CB transition $\ensuremath{\boldsymbol{d}}^{(1)}
  (t)$. (a) and (b) show the $s$-and $p$-polarized THz yields
  $S_{\ensuremath{\operatorname{THz}}}^s (\tau, \theta)$ and
  $S_{\ensuremath{\operatorname{THz}}}^p (\tau, \theta)$, respectively.}
\end{figure}

\section{Photocurrent and Single-Atom Photocurrent Model}

In Eq. (5) in the main text, neglecting the neutral depletion is referred to
as the single-atom photocurrent (SPC) model. The $S_{\text{THz}, s (p)}$ from
the SPC model is shown in Fig. 1(c) of the main text. Here, the
$S_{\text{THz}, s (p)}$ from the traditional PC model is presented in Fig. S5
by comparison. It is shown that there is no significant deviation of the SPC
from the PC which involves the neutral depletion.

\begin{figure}[h]
  \includegraphics{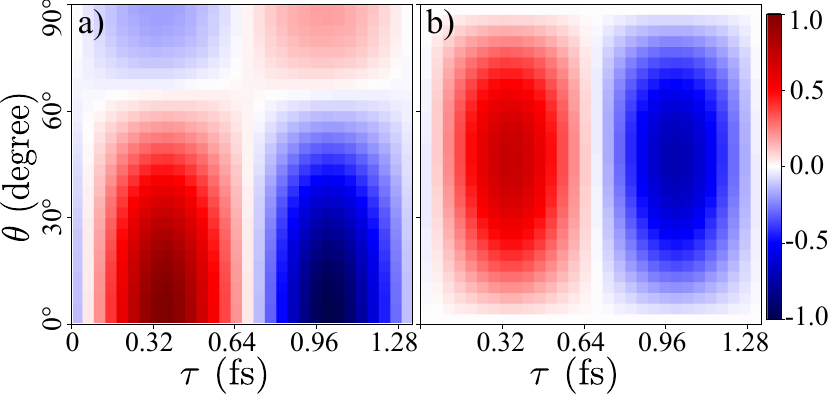}
  \caption{The $S_{\ensuremath{\operatorname{THz}}, s} (\tau, \theta)$ and
  $S_{\ensuremath{\operatorname{THz}}, p} (\tau, \theta)$ evaluated from the
  PC model.}
\end{figure}

For more detailed comparison, slicing the data $S_{\text{THz}, s}$ as
presented by Figs. (a1), (b1) and (c1) in the main text, the data
$S_{\ensuremath{\operatorname{THz}}, s}  (\tau = 0.33 \text{ fs}, \theta)$ and
$S_{\ensuremath{\operatorname{THz}}, s}  (\tau, \theta = 90)$ from experiment,
PC and SPC models are shown in Fig. S6.

\begin{figure}[h]
  \includegraphics{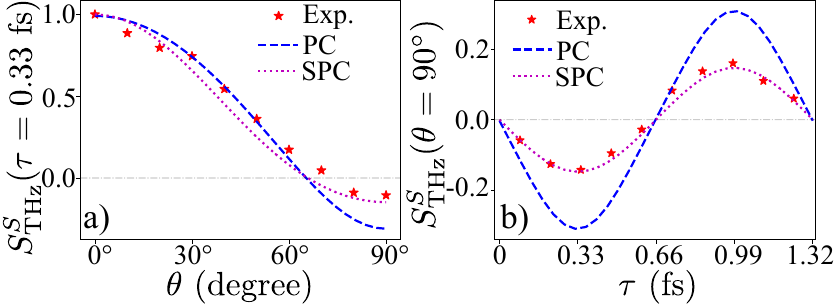}
  \caption{(a) The $S_{\ensuremath{\operatorname{THz}}, s}  (\tau = 0.33
  \text{ fs}, \theta)$ and (b) $E_{\ensuremath{\operatorname{THz}}, s}  (\tau,
  \theta = 90)$ of the experiment (red star), PC (blue dashed) and SPC
  (magenta dotted) models.}
\end{figure}


\end{document}